\documentclass[]{emulateapj}

\shortauthors{Bean et al.}

\begin{document}

\title{Ground-Based Transit Spectroscopy of the Hot-Jupiter WASP-19b in the Near-Infrared}

\author{
Jacob L.~Bean\altaffilmark{1,7},
Jean-Michel D{\'e}sert\altaffilmark{2,8},
Andreas Seifahrt\altaffilmark{1},
Nikku Madhusudhan\altaffilmark{3},
Igor Chilingarian\altaffilmark{4,5},
Derek Homeier\altaffilmark{6}, \&
Andrew Szentgyorgyi\altaffilmark{4}
}

\email{E-mail: jbean@oddjob.uchicago.edu}

\altaffiltext{1}{Department of Astronomy and Astrophysics, University of Chicago, 5640 S.~Ellis Ave, Chicago, IL 60637, USA}
\altaffiltext{2}{Division of Geological and Planetary Sciences, California Institute of Technology, MC 170-25 1200, E.~California Blvd., Pasadena, CA 91125, USA}
\altaffiltext{3}{Department of Physics \& Department of Astronomy, Yale University, P.O.~Box 208120, New Haven, CT 06520, USA}
\altaffiltext{4}{Smithsonian Astrophysical Observatory, 60 Garden Street, Cambridge, MA 02138, USA}
\altaffiltext{5}{Sternberg Astronomical Institute, Moscow State University,
13 Universitetsky prospect, 119992 Moscow, Russia}
\altaffiltext{6}{Centre de Recherche Astrophysique de Lyon, UMR 5574, CNRS, Universit\'e de Lyon, \'Ecole Normale Sup\'erieure de Lyon, 46 All\'ee d'Italie, F-69364 Lyon Cedex 07, France}
\altaffiltext{7}{Alfred P.~Sloan Research Fellow}
\altaffiltext{8}{Sagan Fellow}

\begin{abstract}
We present ground-based measurements of the transmission and emission spectra of the hot-Jupiter WASP-19b in nine spectroscopic channels from 1.25 to 2.35\,$\mu$m. The measurements are based on the combined analysis of time-series spectroscopy obtained during two complete transits and two complete secondary eclipses of the planet. The observations were performed with the MMIRS instrument on the Magellan\,II telescope using the technique of multi-object spectroscopy with wide slits. We compare the transmission and emission data to theoretical models to constrain the composition and thermal structure of the planet's atmosphere. Our measured transmission spectrum exhibits a scatter that corresponds to 1.3 scale heights of the planet's atmosphere, which is consistent with the size of spectral features predicted by theoretical models for a clear atmosphere. We detected the secondary eclipses of the planet at significances ranging from 2.2 to 14.4$\sigma$. The secondary eclipse depths, and the significances of the detections increase towards longer wavelengths. Our measured emission spectrum is consistent with a 2250\,K effectively isothermal 1-D model for the planet's dayside atmosphere. This model also matches previously published photometric measurements from the \textit{Spitzer Space Telescope} and ground-based telescopes. These results demonstrate the important role that ground-based observations using multi-object spectroscopy can play in constraining the properties of exoplanet atmospheres, and they also emphasize the need for high-precision measurements based on observations of multiple transits and eclipses.
\end{abstract}


\section{INTRODUCTION}
Ground-based observations using transit techniques have been playing an increasingly important role in the study of exoplanetary atmospheres over the last few years. It wasn't that long ago that the first ground-based detections of the transmission \citep{redfield08, snellen08} and thermal emission \citep{demooij09, sing09, gillon09} of exoplanets were obtained. Yet today, ground-based transit measurements using photometric techniques are now standard. The photometric technique has seen its widest application to measurements of thermal emission from hot-Jupiters \citep{rogers09, croll10a, croll10b, anderson10, gibson10, alonso10, lopezmorales10, croll11b, demooij11, smith11, burton12, zhao12a, crossfield12b, zhao12b, deming12, lendl13}, but it has also been used for measurements of transmission spectrum of a cool super-Earth \citep{croll11a, demooij12, murgas12, narita12}. 

\begin{deluxetable*}{ccccccl}
\tabletypesize{\scriptsize}
\tablecolumns{7}
\tablewidth{0pc}
\tablecaption{Observing Log}
\tablehead{
 \colhead{UT Date} &
 \colhead{Event} &
 \multicolumn{2}{c}{Exposures} &
 \colhead{Airmass} &
 \colhead{Seeing} &
 \colhead{Conditions}\\
 \colhead{} &
 \colhead{} &
 \colhead{Time (s)} &
 \colhead{\#} &
 \colhead{} &
 \colhead{} &
 \colhead{}\\
}
\startdata
2012 Mar 11 00:18 $\rightarrow$ 2012 Mar 11 06:44 & Eclipse & 40 & 422 & 1.29 $\rightarrow$ 1.04 $\rightarrow$ 1.37 & 0.7\arcsec & clear\\
2012 Mar 13 00:01 $\rightarrow$ 2012 Mar 13 03:55 & Transit & 40, 30 & 304 & 1.32 $\rightarrow$ 1.04 $\rightarrow$ 1.06 & 0.5\arcsec & clear \\
2012 Apr 04 00:42 $\rightarrow$ 2012 Apr 04 06:47 & Transit & 40 & 408 & 1.07 $\rightarrow$ 1.04 $\rightarrow$ 2.00 & 0.8\arcsec & occasional thin cirrus \\
2012 Apr 05 23:23 $\rightarrow$ 2012 Apr 06 05:49 & Eclipse & 40 & 430 & 1.17 $\rightarrow$ 1.04 $\rightarrow$ 1.60 & 0.7\arcsec & mostly clear, passing\\
 & & & & & & cirrus for 15\,min
\enddata
\label{tab:log}
\end{deluxetable*}		


While photometry is useful, high-precision spectroscopy over wide bandpasses is ultimately needed to significantly improve our understanding of exoplanet atmospheres. Only these kind of data can unambiguously reveal atmospheric properties by resolving spectral features and distinguishing the overlapping lines from different chemical species. Interpreting sparse photometry requires heavy reliance on theoretical models that have to make assumptions about many of the fundamental properties of the planets that we ultimately want to determine observationally. Furthermore, spectroscopy opens up the possibility of discovering unanticipated phenomena much more so than photometry. 

The development of ground-based transit spectroscopy of exoplanet atmospheres has not enjoyed parallel maturation with the photometric technique, and such measurements are still very rare. There are two reasons for this. One reason is that ground-based observations using photometric techniques can acquire simultaneous observations of reference stars to enable high-precision relative corrections for variations in Earth's atmospheric transparency, while this is not possible with single-object spectrographs. The second reason is that spectrographs typically have input feeds, either slits or fibers, that do not encircle all the light from a source. This introduces random variations in the light measured at the detector due to variations in seeing and imperfect guiding. While there has been some development of post-processing algorithms to correct single-object, narrow-slit transit spectroscopy for variations in atmospheric transparency and slit losses \citep{swain10, crossfield11, crossfield12a, waldmann12}, the technique has not been widely adopted.

We have recently developed a technique for ground-based transit spectroscopy of exoplanet atmospheres that overcomes the limitations of single-object, narrow-slit observations. The idea is to use multi-object slit spectrographs to obtain simultaneous observations of a target and reference stars so that differential spectroscopy analogous to differential photometry can be performed. Light losses in the image plane are eliminated by using wide slits ($>$\,10\arcsec\ is recommended). We have previously used this new technique to place constraints on the transmission spectrum of the super-Earth GJ\,1214b in the optical and near-infrared \citep{bean10, bean11}. \citet{gibson13} have also recently used this technique to study the optical transmission spectrum of the hot-Jupiter WASP-29.


We present in this paper transit and secondary eclipse observations of the ultra short-period ($P$\,=\,0.8\,d) hot-Jupiter WASP-19b \citep{hebb10}. As a low-density hot-Jupiter ($R_{p}\,=\,1.31\,R_{Jup}$, $M_{p}\,=\,1.15\,M_{Jup}$) orbiting very close to a star slightly cooler and smaller than the Sun ($R_{\star}\,=\,0.94\,R_{\odot}$), WASP-19b is expected to have some of the largest spectral features in transmission and emission among the known transiting exoplanets. WASP-19 itself is also is an intermediate-brightness ($H$\,=\,10.6) transiting planet host star. This means that there are numerous similar brightness reference stars in the roughly 5$\arcmin$ field-of-view typical of multi-object slit spectrographs.

Observations of the transmission and emission spectrum of WASP-19b have the potential to address numerous questions that have arisen concerning the nature of hot-Jupiter atmospheres. Our motivation was to obtain data that could be used to determine the composition and temperature structure of the planet's atmosphere. Such information would, for example, be useful for constraining theories about the origin of thermal inversions \citep[e.g.,][]{hubeny03, burrows07, fortney08, knutson10}, the statistics of albedo and heat redistribution \citep[e.g.,][]{burrows08, budaj11, cowan11}, and the possibility of high carbon-to-oxygen abundance ratios (C/O) in hot-Jupiter atmospheres \citep[e.g.,][]{madhusudhan12}.

The paper is laid out as follows. We present the new data we have obtained for WASP-19b in \S2. We describe the analysis of the transit and eclipse light curves to determine the transmission and emission spectrum of the planet in \S3. The implications of these data for the properties of WASP-19b's atmosphere are considered in \S4. We conclude in \S5 with a discussion of the results.

\section{DATA}
\subsection{Observations}
We obtained time-series spectroscopy during two transits and two secondary eclipses of the planet WASP-19b using the MMIRS instrument \citep{mcleod12} on the Magellan II (Clay) telescope at Las Campanas Observatory in March and April 2012. A log of the observations is given in Table~\ref{tab:log}. We used the multi-object mode of MMIRS with a slit mask that allowed us to gather spectra simultaneously for WASP-19 and three other stars. As for our previous observations using this technique \citep{bean10,bean11}, the slit widths were 12\arcsec\ to avoid slit losses. The slit lengths were 30\arcsec, which provided significant coverage of the sky background. We used an $HK$ grism as the dispersive element and an $HK$ filter to isolate the first order spectra. Spectra from 1.22 to 2.36\,$\mu$m with a dispersion of 6.6\,\AA\ pixel$^{-1}$ were obtained for all the objects.

Complete transits and eclipses were observed without interruption, and a minimum of 30 minutes of data were obtained in each case both before ingress and after egress. All data were obtained at airmasses less than 2.0, with most obtained at airmasses less than 1.5. The exposure times used were 30 and 40\,s, and the overhead was typically 14\,s including the time to read and reset the detector and write the data to disk. The off-axis guide camera and wavefront sensor associated with the instrument were used to maintain stable pointing and the correct shape of the primary mirror continuously during the observations. 

Although the wavelength range we observed in contains many narrow sky emission lines, we did not nod the telescope during the observations to enable pairwise image subtraction for removing the background. This staring approach is justified because the edges of the slits in the mask are likely reasonably parallel, and thus the sky lines have the same instrumental profile in the spatial direction. Furthermore, the sky emission dominates the background over the telescope emission at these near-infrared wavelengths, and thus the spatial variability can not be removed any better by nodding than simply interpolating over the spatially resolved background spectra obtained in each exposure. The staring approach is advantageous for high-precision time-series spectroscopy not only because it saves overhead associated with periodically moving the telescope, but also because it keeps the data on the same pixels throughout, and therefore minimizes the influence of imperfect flat fielding.

The observations were obtained in generally clear conditions. There were occasional thin cirrus clouds for two of the nights, and drops in the fluxes can be seen in the raw spectrophotometry corresponding to the passing clouds. However, these drops in fluxes are similar for all the stars, and the WASP-19 light curves are well-corrected for this effect using the reference star data.

\subsection{Data reduction}
We reduced the data using a similar approach as for our previous MMIRS observations \citep{bean11} with two deviations. The steps in the process included collapsing the sample-up-the-ramp data cubes in to a single frame of total counts for the individual exposures, applying flat field corrections based on spectroscopic flats taken with an internal calibration lamp, subtraction of the background on a wavelength-by-wavelength basis for each spectrum, extraction of the spectra, and determination of the wavelength solution based on an arc lamp exposure taken using a mask having 0.5\arcsec\ wide slits. The deviations from our previously used data reduction algorithm are (1) ignoring the first non-destructive read in the up-the-ramp samples, and (2) testing the application of non-linearity corrections. We discuss these two issues in the following.

\subsubsection{Fitting the up-the-ramp samples}
The data were obtained in the ``up-the-ramp'' sampling mode with 5\,s per read. In this mode, which is only possible for hybrid CMOS arrays, the detector is read out non-destructively multiple times during an exposure. The MMIRS system saves these images to disk, and so each exposure involves a sequence of full frame images. This sampling technique yields measurements of the flux as it accumulates in the pixels before they are reset to begin another exposure. In the limit of a constant illumination and a linear pixel response, the flux values for a pixel will exhibit a constant rise with time over the course of an exposure. The advantage of this technique is that it can reduce the per-exposure read noise, and it enables checks for detector systematics.

The first step in our data reduction is to collapse the corresponding data cubes of non-destructive reads to a single frame with the total counts per pixel for the exposure. This is done by fitting a linear trend to the ramp samples as a function of elapsed time on a pixel-by-pixel basis, and then calculating the total counts recorded for the pixels by multiplying the slope of this trend by the exposure time. 

\begin{figure}
\resizebox{\hsize}{!}{\includegraphics{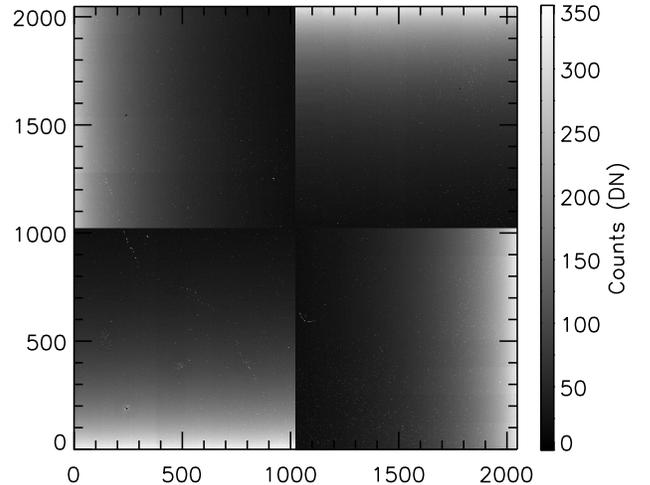}}
\caption{Image of the difference between the first and second non-destructive reads in a dark exposure.}
\label{fig:dark}
\end{figure}

During the analysis of the data, we noticed that the first read, which corresponds to a zero second integration, always has a significantly lower bias level than the subsequent reads in a given exposure. Figure~\ref{fig:dark} shows the difference between the first two 5\,s non-destructive reads in a dark image that illustrates this effect. Differences of up to and beyond 350\,DN are observed, and the magnitude of the difference smoothly increases towards the ends of the readout channels. The dark current for the MMIRS detector system is estimated to be approximately 0.01\,DN\,s$^{-1}$\,pixel$^{-1}$, and is a negligible contributor to the observed difference between the first two ramp samples. This ``bias drift'' after reseting the detector is also exhibited by the HAWAII-2\,RG chip in the similar MOSFIRE instrument \citep{mclean10}, although the magnitude of the effect seems to be significantly larger in the MMIRS detector \citep{kulas12}.

The HAWAII-2 chip used in MMIRS does not have light-insensitive reference pixels like the later generation HAWAII-2\,RG chips (the ``R'' in the RG stands for reference pixels). Therefore, there is no straightforward way to calibrate the MMIRS data for this effect using external information. By examining dark frames, we noticed that the bias drift is only significant between the first and second non-destructive reads in dark frames, and the bias level is constant within the expected noise level for subsequent reads. Under the assumption that whatever process in the detector electronics is causing this effect settles after $<$\,5\,s, we elected to ignore the first non-destructive read when fitting the up-the-ramp samples. There is enough information with the other six or eight non-destructive reads per 30 or 40\,s exposure, respectively, to simultaneously fit for an offset corresponding to the bias level and a slope corresponding to the count rate. Therefore, the final total counts recorded for a pixel is still the slope of the fit as a function of read time multiplied by the exposure time, but the zero point of the fit is not fixed to a known bias level like it otherwise could be.

In our previous analysis of MMIRS data \citep{bean11}, we noticed a systematic noise pattern in the up-the-ramp samples in one of our data sets. The effect was severe enough to prompt us to consider those particular data unreliable, and to ignore them in our analysis. We note that we did not observe this same effect in the current data sets. All the obtained data are included in our analysis.

\begin{figure}
\resizebox{\hsize}{!}{\includegraphics{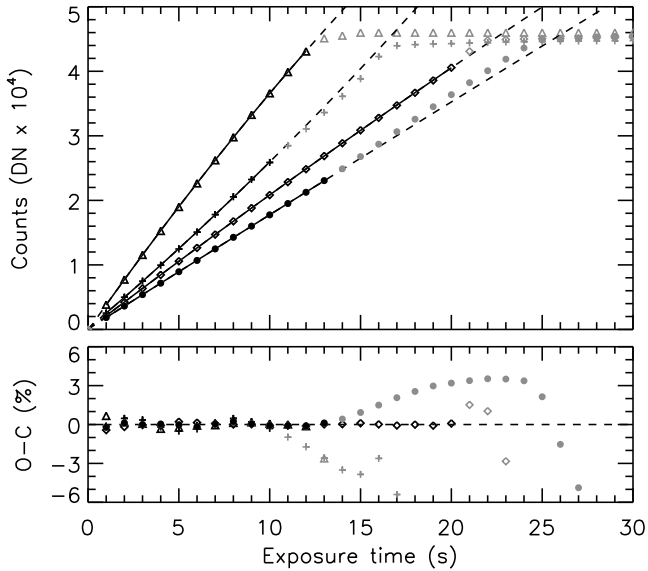}}
\caption{\textit{Top} Counts in DN for non-destructive reads in a flat field image taken to investigate the non-linear behavior of the detector. The different point styles represent examples from the different quadrants of the detector. The four examples have different slopes (count rates) because there is a strong instrumental throughput gradient across the field of view. The lines are fits to the data using a quadratic polynomial as a function of time (Eq.\ 1). The solid lines are the fits over the ranges where this functional form provides a good description of the data. The upper limits of these ranges correspond to the maximum values that can be corrected to high accuracy based these low-order fits. The data points are black in the well-fit regime, and grey outside. The dashed lines show the extrapolation of the fits. \textit{Bottom} Residuals from the fit in terms of percentage of the data values. The point styles are the same as for the top panel.}
\label{fig:nonlin}
\end{figure}

\subsubsection{Detector non-linearity}
Another deviation from our previous analysis of MMIRS time-series spectroscopy was to consider the influence of the non-linear response of the detector. We studied the non-linear behavior of the MMIRS detector by examining a series of flat-field exposures taken with the instrument in imaging mode using the internal calibration lamp and an $H$ filter. As is typical of near-infrared detectors, the MMIRS chip displays significant non-linear response well before the full-well depth of the pixels is reached. The full-well is approximately 45,000\,DN (for a gain of 5\,$e^{-}$\,DN$^{-1}$), and only 61\% of the pixels are linear within 1\% up to 16,000\,DN. Furthermore, each of the four quadrants in the detector exhibits different behavior. Two quadrants even display a faster, rather than the expected slower, than linear rise in counts before turning over to plateau near saturation. The pixel-to-pixel variations in the non-linear behavior within a quadrant are also significant.

We were aware of the approximately 16,000\,DN transition to non-linearity based on our previous observations, and so the exposure times of the WASP-19 data were tuned to keep the peak counts approximately at or below this level. Nevertheless, a significant number of pixels exhibit $>$\,1\% non-linearity even below this limit, and a few pixels near the peak in flux for the $H$-band region of the spectra reach upwards of 20,000\,DN in some exposures that were taken during periods of better than average seeing and low airmass. This motivated us to derive non-linearity corrections and apply them to the data to test the influence of this effect on our results. 

We determined non-linearity corrections based on a quadratic polynomial function. Coefficients for the corrections were derived by fitting the flux as a function of time for the non-destructive reads in flat field frames obtained using a total exposure time of 50\,s with 1\,s up-the-ramp samples. This was done for each pixel separately. The fitted function was
\begin{equation}
F = a + bt + ct^2,
\label{eq1}
\end{equation}
where $F$ are the fluxes of the non-destructive reads, $t$ are the corresponding elapsed times, and $a$, $b$, and $c$ are the parameters in the fit. The formula for correcting the measured fluxes is then 
\begin{equation}
F' = \frac{-b^2 + b^2\sqrt{1 - \frac{4c(a - F)}{b^2}}}{2c},
\label{eq2}
\end{equation}
where $F'$ is the corrected flux.

The coefficients were determined by fitting the data after subtracting the first non-destructive read to approximately remove the bias. This was done because the non-linearity is related to the process of electron creation in the photosensitive layer, and this isn't influenced by the bias level, which arises during the reset of the detector. In determining the coefficients, the first non-destructive read wasn't fit just as for the science data, and the intercept term of the quadratic polynomial was left as a free parameter. 

We iteratively fit the data considering a range of possible upper limits in count levels to determine the range over which the corrections were accurate using our low-order functional form. The median value for the upper limits across the detector is 32,000\,DN. However, a small fraction of pixels are not well-described by a quadratic polynomial much beyond their 1\% linearity ranges. Examples of the non-linearity fits to some of the data are shown in Figure~\ref{fig:nonlin}. The determined polynomial coefficients were consistent for 50 separate exposures. The averages of the determined coefficients among these 50 frames were adopted as the final correction coefficients.

We applied the non-linearity corrections to the values for the non-destructive reads in each exposure before fitting for the count rate. The first non-destructive read was subtracted before applying the corrections to be consistent with the convention for the determination of the coefficients. This read was then ignored in the fit of the linear trend to determine the non-linear-corrected count rate. 

The results for the final light curves (derived transit parameters and model fit residuals) are not significantly different ($\ll\,1\sigma$) when using or not using the non-linearity corrections. This is because most of the pixels in the WASP-19 data sets had counts below the 1\% non-linearity level. For the results presented here, we did not apply the non-linearity corrections in order to avoid unnecessary processing steps that could add noise. The corrections would in principle yield more accurate results for more heavily exposed data, and thus could be of use to the wider community. Tables of the coefficients needed to apply Equation\,\ref{eq2} and the applicable ranges of the corrections are available from us upon request.

\subsection{Creation of the light curves}
\begin{figure}
\resizebox{\hsize}{!}{\includegraphics{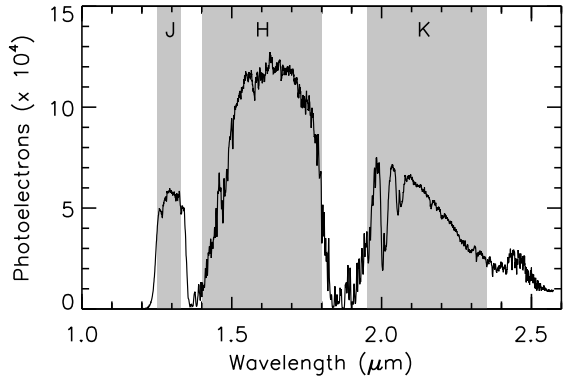}}
\caption{An example spectrum extracted from the MMIRS data for WASP-19.}
\label{fig:star_spectrum}
\end{figure}

An example extracted spectrum for WASP-19 from a 40\,s exposure is shown in Figure~\ref{fig:star_spectrum}. This spectrum has a signal-to-noise ratio of 200\,pixel$^{-1}$ at the peak of the flux at 1.6\,$\mu$m. We created spectrophotometric light curves for WASP-19 and the reference stars by summing the extracted spectra over wavelength. We created one, four, and four channels of light curves from the spectra in the $J$-, $H$-, and $K$-band atmospheric windows, respectively. The limits of the bandpasses in each of the atmospheric windows are illustrated in Figure~\ref{fig:star_spectrum}.

\begin{figure*}
\resizebox{\hsize}{!}{\includegraphics{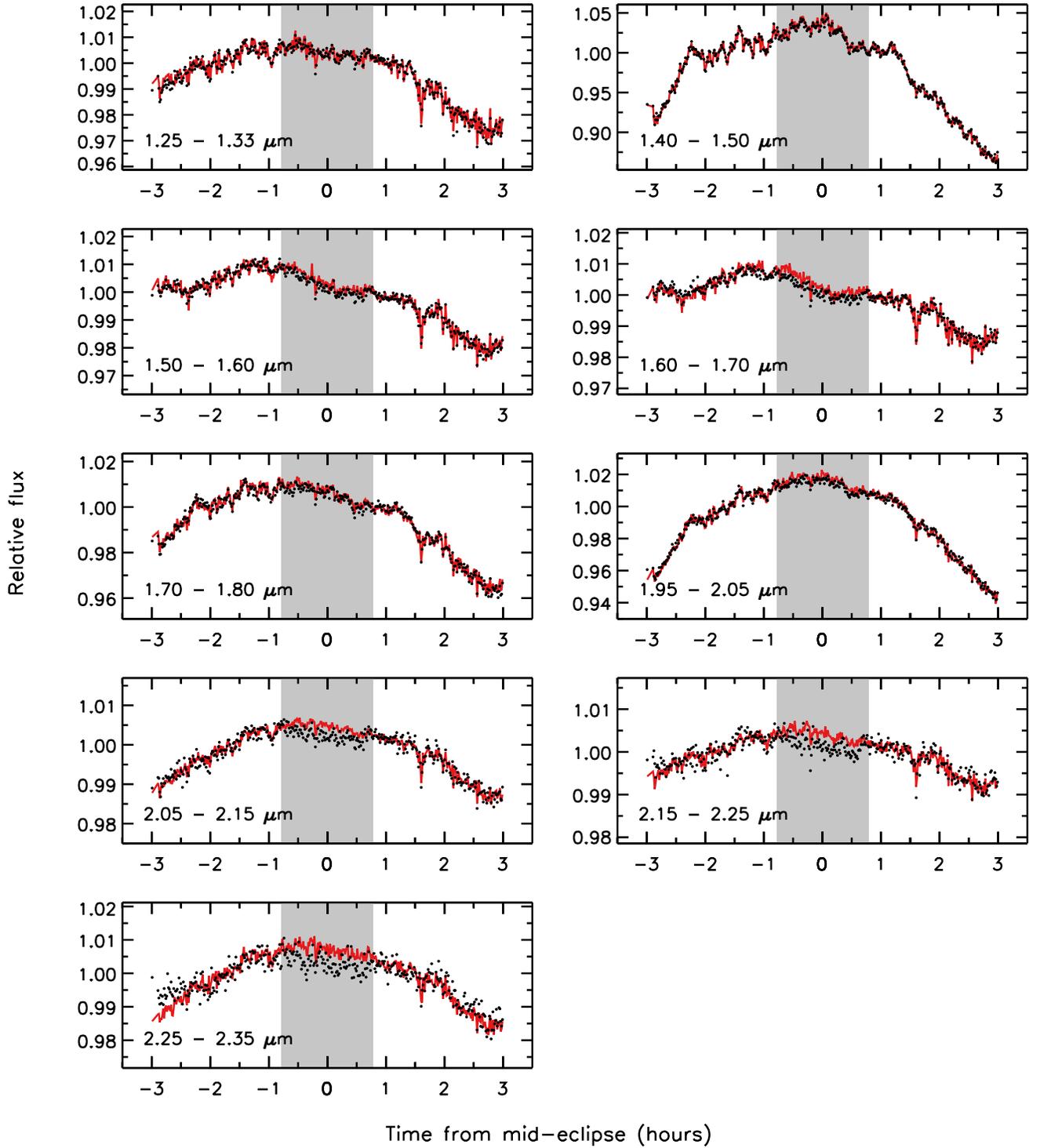}}
\caption{Raw spectrophotometric light curves of WASP-19 (circles) and the corresponding correction functions (red lines) for the secondary eclipse observations obtained on UT 2012 April 5/6. The grey shaded region indicates the time where the planet is occulted (first to fourth contact).}
\label{fig:decor_lc}
\end{figure*}

We divided the time-series spectrophotometric light curves for WASP-19 by the sum of reference star data to correct for the variability of Earth's atmospheric transparency during the observations. The choice of which reference stars to use for this correction was guided by which combination gave the smallest residuals in the final light curve fits. We found that only using the reference star closest on the sky to WASP-19 yielded the smallest model fit residuals for the $J$- and $H$-band data, while the sum of all three observed reference stars gave the best results for the $K$-band data. This same result was obtained for each of the four nights of data independently.

The light curves for WASP-19 after correction with the reference star data exhibit a slow, smooth trend with time similar to trends that are common in ground-based near-infrared transit photometry \citep[e.g.,][]{croll10a,croll10b,croll11a,croll11b,sada10}. We modeled this trend as a second order polynomial simultaneously with the light curve fitting (see \S\,3). The derived trend with time was not the same for the different spectral channels for a given event.

We searched for additional and alternative decorrelation functions with respect to airmass and pixel position, but we did not detect statistically significant correlations with these parameters. This is in contrast to our previous analysis of MMIRS transit spectroscopy \citep{bean11}. The lack of a strong correlation with airmass could be explained by the more similar color of WASP-19 with the reference stars compared to the target of our previous study, the mid-M dwarf GJ\,1214, and its corresponding reference stars. The stability of the spectral positions was slightly better for the current data sets, but not significantly so. Therefore, this likely is not the reason for the lack of a correlation with spatial pixel position compared to the previous study. One possible explanation for the difference is that the seeing was typically worse by factors of two to three for the WASP-19 data compared to the previously analyzed data. This means that the WASP-19 data were spread over more pixels (spatial profile FWHMs of 6 to 10 pixels compared to 3 pixels), and this could have mitigated the influence of pixel-to-pixel sensitivity variations not corrected by the flat fielding.

Because the light curves do not require decorrelation against instrument-related variables, there is little gain in considering data obtained significantly before or after the transits and eclipses. Indeed, the longer the extent of the time-series, the more likely it is that the atmospheric effect leading to the slow, smooth trend with time will not be well-described by a low-order polynomial. Therefore, we limited our analyses to the data taken within three hours of mid-transit/eclipse. The trimmed data still encompass more than two hours out-of-transit/eclipse data before and after in the cases that the observations spanned this length of time.

Raw light curves for the April 5/6 secondary eclipse data set are shown in Figure~\ref{fig:decor_lc}. The data for WASP-19 are shown in this figure along with the corresponding correction functions. The correction functions are a combination of the reference star data, which captures the dominant variability, and the decorrelation against time. An example of the effect of passing cirrus can be seen by the short drop in flux about 1.5\,hours after mid-eclipse that is common between WASP-19 and the reference stars.

\begin{deluxetable}{lrcl}
\tabletypesize{\scriptsize}
\tablecolumns{4}
\tablewidth{0pc}
\tablecaption{Transit Parameters}
\tablehead{
 \colhead{Parameter} &
 \multicolumn{3}{c}{Value}
}
\startdata
$i$ (\degr)           &   78.73 & $\pm$ & 0.20 \\
$b$                   &   0.681 & $\pm$ & 0.008 \\
$P$ (d)               &   0.78883910 & $\pm$ & 0.00000011 \\
T$_{c1}$ (BJD$_{TBD}$) &   2455512.90256 & $\pm$ & 0.00007
\enddata
\label{tab:transit_param}
\end{deluxetable}

\begin{deluxetable}{cc}
\tabletypesize{\scriptsize}
\tablecolumns{2}
\tablewidth{0pc}
\tablecaption{Event Times}
\tablehead{
 \colhead{Mid-Event Time (BJD$_{\mathrm{TDB}}$)\tablenotemark{a}} &
 \colhead{O - C (s)}
}
\startdata
Transits \\[0.5mm]
2455999.616301\,$\pm$\,7.0E-5 & 1.8\,$\pm$\,6.0\tablenotemark{b} \\
2456021.703740\,$\pm$\,8.5E-5 & -3.0\,$\pm$\,7.3\tablenotemark{b} \\[2mm]
\hline\\
Eclipses \\[0.5mm]
2455997.6440\,$\pm$\,1.0E-3 & -28\,$\pm$\,86 \tablenotemark{c}\\
2456023.6763\,$\pm$\,1.0E-3 & +21\,$\pm$\,89 \tablenotemark{c}
\enddata
\tablenotetext{a}{BJD$_{\mathrm{TDB}}$ is the Barycentric Julian Date in the Barycentric Dynamical Time standard \citep{eastman10}.}
\tablenotetext{b}{Residuals from the ephemeris given in Table\,\ref{tab:transit_param}.}
\tablenotetext{c}{Residuals from the prediction of when the orbital phase is 0.5 assuming a circular orbit and the ephemeris given in Table\,\ref{tab:transit_param}, and including the light travel time effect.}
\label{tab:times}
\end{deluxetable}		

\subsection{Properties of the light curves}
The residuals from the model fits to the light curves when each of the two transit and secondary eclipse time-series are binned together at a sampling of one minute have rms values ranging from 650 to 1608\,ppm. Most of the combined light curves have residuals smaller than 1000\,ppm. The residuals are 2.0 to 3.1 times the expected level given the uncertainties estimated during the reduction process (including the noise contribution from the reference star division). In our previous analysis of MMIRS data, we noted that the $H$-band data had significantly larger resdiuals than expected relative to the $J$- and $K$-band data. We do not find such an effect in the current data. Instead, the different spectrophotometric channels all have a similar level of noise relative to what is expected.

\section{ANALYSIS}
We fitted the spectrophotometric light curves for WASP-19 with transit and eclipse models multiplied by normalization and decorrelation (the quadratic function of time described in \S\,2.3) functions. We used the exact analytic formulae given by \citet{mandel02} for the transit and eclipse models. The parameterizations for these models are described in the subsections below. We assumed the planet is in a circular orbit, which is supported by the existing radial velocity data and secondary eclipse times \citep{anderson10, gibson10, hellier11, burton12, anderson13}. The orbital period of the planet (P) was fixed to the value determined from the transit times as described below. The normalization and decorrelation functions added an additional three parameters for each light curve. 

We identified the best-fit models and corresponding parameters for the analyses described below using a non-linear least squares algorithm \citep{markwardt09}. We determined the confidence intervals on the best-fit parameters using a residual permutation bootstrap (``prayer bead'') algorithm. The uncertainties in the light curve points estimated during the reduction process were scaled to give a reduced $\chi^{2}$\,=\,1 for the best-fit model. The scaling factors range from 2.0 to 2.9.  The uncertainties derived from the residual permutation bootstrap range from 1.0 to 2.5 times larger than the errors derived from a Markov Chain Monte Carlo technique, which suggests that correlated noise is the limiting factor for the observations.

\subsection{Transits}

\subsubsection{Model parameterization}
We parameterized the model for the transits using the square of the planet-to-star radius ratio ((R$_{p}$/R$_{\star}$)$^{2}$), the orbital inclination of the planet ($i$), the impact parameter ($b$\,$\equiv$\,$a$\,/\,R$_{\star}$\,$cos\,i$), quadratic limb darkening coefficients ($\gamma_{1}$ and $\gamma_{2}$), and the central transit times (T$_{c1}$). We used the same values for $i$ and $b$ for all the light curves. We assumed that the transit times were the same for all the spectrophotometric channels for a given event. The (R$_{p}$/R$_{\star}$)$^{2}$ values were determined for each spectrophotometric channel, and were assumed to be same for a given wavelength for both transit data sets when the two data sets were analyzed together.

\subsubsection{Limb darkening}
Determined transit depths from light curve fitting are strongly correlated with the adopted limb darkening description even at the near-infrared wavelengths considered in the current study. To bring additional constraints on the light curve fits and potentially increase the sensitivity of our analysis to the transmission spectrum of the planet's atmosphere, we estimated quadratic limb darkening coefficients using model atmospheres computed with the PHOENIX code \citep{hauschildt99} for the stellar parameters (T$_{eff}$\,=\,5500\,K, log g\,=\,4.5, and [M/H]\,=\,0.0) estimated by \citet{hebb10}, which are consistent with the results from \citet{doyle13}. However, the light curve fits were significantly worse when using these coefficients as compared to when using a single linear coefficient determined solely from the data themselves. For example, the models predict $\gamma_{1}$ values ranging from 0.3 to 0.4 when $\gamma_{2}$ is set to zero, while the best-fit values with no constraints are below 0.2 except in the bluest channel (see Table\,\ref{tab:transit_depths}). The disagreement between the theoretical and empirical estimates was not ameliorated when using model stellar atmospheres with the WASP-19's nominal T$_{eff}$ increased or decreased by 150\,K. This is not an unusual situation when modeling high-precision light curves \citep[e.g.,][]{knutson07}, and likely stems from the inadequacy of 1D hydrostatic models to accurately represent stellar atmospheres \citep{hayek12}. For all analyses presented in this paper we allow 
$\gamma_{1}$ to vary with no outside constraints, and we fix $\gamma_{2}$\,=\,0 because our tests indicated that the data do not warrant an additional parameter.

\begin{figure*}
\resizebox{\hsize}{!}{\includegraphics{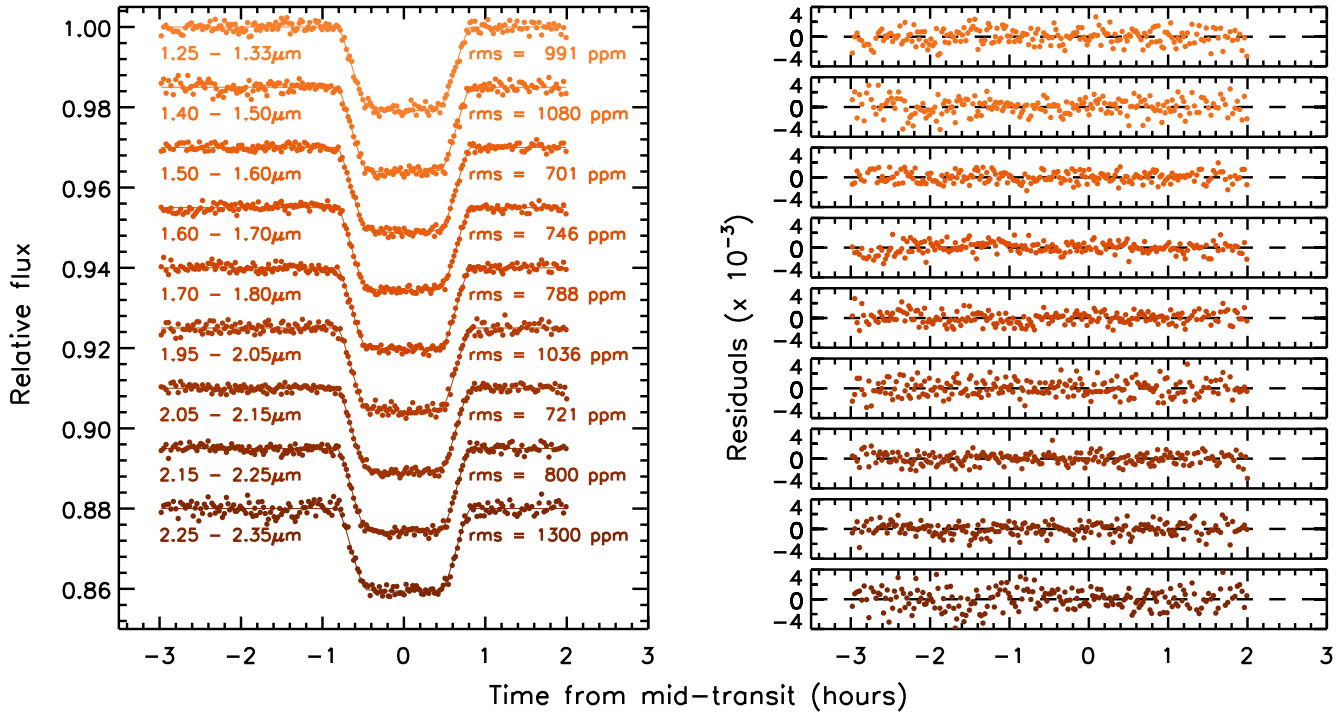}}
\caption{\textit{Left} Transit light curves (circles) and best-fit models (lines) for the WASP-19b MMIRS data. The data for the two transits have been combined and binned to a sampling of one minute. \textit{Right} Residuals from the best-fit models.}
\label{fig:transit_lc}
\end{figure*}

\subsubsection{Determining the transit parameters}
The first step in our analysis of the transit light curves was to estimate improved parameters for the system because the MMIRS data are significantly more precise than existing data. We began by fitting the two transit events separately with all the parameters free. The determined parameters were consistent between the two data sets. We then performed a combined analysis of the data sets requiring the parameters to be consistent for the events as described above. The determined physical values for the transit in this case were also consistent with the most recently reported values in the literature \citep{anderson13}. To take advantage of the constraints offered by the previously obtained data, we included priors for our final fit on $i$ (79.42\degr\,$\pm$\,0.39\degr) and $b$ (0.656\,$\pm$\,0.015) based on the determined parameters in \citet{anderson13}. 

The individual transit times determined by fitting the MMIRS data are given in Table\,\ref{tab:times}. The determined transit parameters, and a new ephemeris based on fitting the new transit times along with the previously published times from \citet{hebb10} and \citet{hellier11} are given in Table\,\ref{tab:transit_param}. Note that we do not use the additional transit time for WASP-19b available in the literature given by \citet{dragomir11} for determining the revised ephemeris because it has a relatively low precision (51\,s). None of the four considered transit times deviates from the new ephemeris by more than 1$\sigma$. The average determined (R$_{p}$/R$_{\star}$)$^{2}$\,=\,0.0207 for the MMIRS data.

\subsubsection{Determining the transmission spectrum}
When assessing a planet's transmission spectrum from transit light curves, the relative depths are typically the key parameters to determine rather than the absolute depths because the pressure level probed by the data is often unknown, and thus comparison of the data to theoretical models for the planet's atmosphere must necessarily include a free offset in the overall level. We re-fit the MMIRS transit light curves after refining the transit parameters with $i$ and $b$ fixed to precisely estimate confidence intervals on the relative transit depths. We also fixed the transit times to the prediction of the ephemeris given in Table\,\ref{tab:transit_param} because there is no evidence of transit timing variations in this system. The determined transit depths and limb darkening coefficients from this analysis are given in Table\,\ref{tab:transit_depths}. The resulting normalized and decorrelated transit light curves with best-fit models and residuals are shown in Figure\,\ref{fig:transit_lc}.

\subsection{Secondary eclipses}
The free parameters for the secondary eclipse models were the planet-to-star flux ratios (F$_{p}$/F$_{\star}$) and central eclipse times (T$_{c2}$). We assumed the planet is a uniform disk. We used the values for $i$ and $b$ (see Table\,\ref{tab:transit_param}) and the average value for the (R$_{p}$/R$_{\star}$)$^{2}$ values (0.0207) determined from the transit modeling. The eclipse times were the same for all the spectrophotometric channels for a given event. The F$_{p}$/F$_{\star}$ values were determined for each spectrophotometric channel.

\begin{deluxetable}{ccc}
\tabletypesize{\scriptsize}
\tablecolumns{3}
\tablewidth{0pc}
\tablecaption{Transit Depths and Limb Darkening Coefficients}
\tablehead{
 \colhead{Wavelength ($\mu$m)} &
 \colhead{(R$_{p}$/R$_{\star}$)$^{2}$\tablenotemark{a}} &
 \colhead{$\gamma_{1}$\tablenotemark{a}}
}
\startdata
  1.25 -- 1.33 & 0.02060 $\pm$ 2.8e-04 & 0.28 $\pm$ 0.06 \\
  1.40 -- 1.50 & 0.02117 $\pm$ 3.1e-04 & 0.15 $\pm$ 0.07 \\
  1.50 -- 1.60 & 0.02106 $\pm$ 1.5e-04 & 0.15 $\pm$ 0.05 \\
  1.60 -- 1.70 & 0.02056 $\pm$ 3.0e-04 & 0.14 $\pm$ 0.05 \\
  1.70 -- 1.80 & 0.02030 $\pm$ 1.6e-04 & 0.12 $\pm$ 0.05 \\
  1.95 -- 2.05 & 0.02060 $\pm$ 3.1e-04 & 0.11 $\pm$ 0.07 \\
  2.05 -- 2.15 & 0.02089 $\pm$ 1.0e-04 & 0.16 $\pm$ 0.03 \\
  2.15 -- 2.25 & 0.02070 $\pm$ 1.8e-04 & 0.09 $\pm$ 0.04 \\
  2.25 -- 2.35 & 0.02048 $\pm$ 4.7e-04 & 0.07 $\pm$ 0.09 \\
\enddata
\tablenotetext{a}{From an analysis with $i$, $b$, and $\gamma_{2}$ fixed to 78.73\degr, 0.681, and 0.0, respectively}
\label{tab:transit_depths}
\end{deluxetable}

\begin{figure*}
\resizebox{\hsize}{!}{\includegraphics{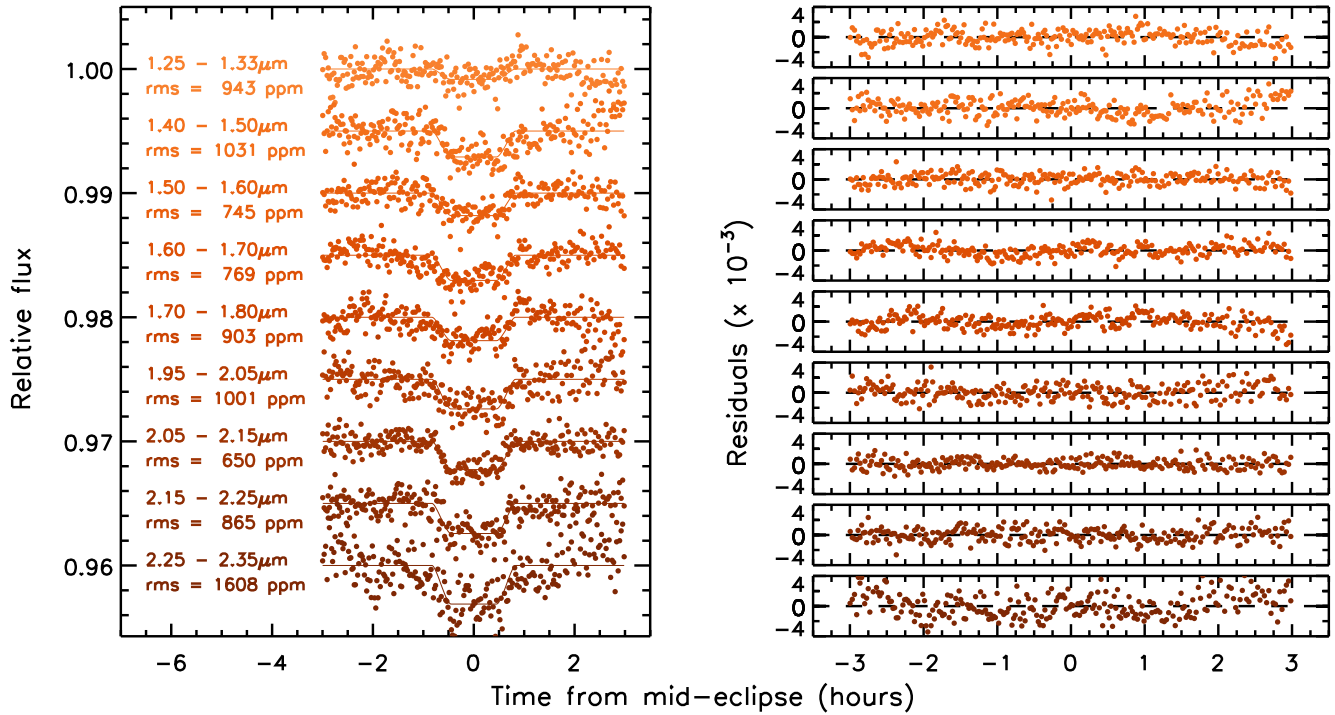}}
\caption{\textit{Left} Secondary eclipse light curves (circles) and best-fit models (lines) for the WASP-19b MMIRS data. The data for the two eclipses have been combined and binned to a sampling of one minute. \textit{Right} Residuals from the best-fit models.}
\label{fig:eclipse_lc}
\end{figure*}

We first fit the light curves for the two secondary eclipses separately to check for systematic errors. The determined F$_{p}$/F$_{\star}$ values for the spectrophotometric channels were consistent between the different events. We then performed a combined analysis of the two data sets with common F$_{p}$/F$_{\star}$ values. The F$_{p}$/F$_{\star}$ values determined from this analysis are given in Table\,\ref{tab:eclipse_depths}, and the determined eclipse times are given in Table\,\ref{tab:times}. The resulting normalized and decorrelated secondary eclipse light curves with best-fit models and residuals are shown in Figure\,\ref{fig:eclipse_lc}.

\subsection{Possible influence of stellar activity}
The star WASP-19 is known to be active \citep{hebb10}, and this could potentially influence our measurements. \citet{tregloan13} have recently reported the detection of star spot crossings in transits of WASP-19b that were observed in 2010. We do not see evidence for spot crossing in our transit light curves. However, the presence of spots on the host star could result in different transit and eclipse depths measured at different epochs even when they are unocculted because the average brightness of the stellar disk might be changing. 

We performed some calculations to determine if our measurements could be significantly influenced by unocculted spots. \citet{hebb10} reported photometric variability of up to 0.8\% over the presumable 10.5\,d rotation period of the star from the WASP-South discovery data \citep[effective central wavelength of approximately 550\,nm,][]{pollacco06}. To estimate the maximum likely effect, we consider the case that the spots leading to the photometric variability observed by \citet{hebb10} are on the Earth-facing side of the star for one of the observations, and either not present or on the opposite side of the star at the other epoch. Applying the formalism from \citet{desert11} \citep[Eq.\ 8 from ][]{berta11}, and assuming the spot and star radiate like blackbodies with a temperature difference of 300\,K and that the surface area fraction covered by spots is 4\%, we find that this situation would result in a 1 x 10$^{-4}$ change in the transit depth at 1.5\,$\mu$m. The same situation would result in a change of 1 x 10$^{-5}$ in the eclipse depths. These values are less than our measurement precisions. Therefore, we conclude that stellar variability likely has not influenced our measurements, and that we can combine our secondary eclipse data with previous data to put joint constraints on the properties of the planet's atmosphere.

\section{CONSTRAINTS ON THE PLANET'S ATMOSPHERE}

\subsection{Transmission Spectrum}
\begin{figure}
\resizebox{\hsize}{!}{\includegraphics{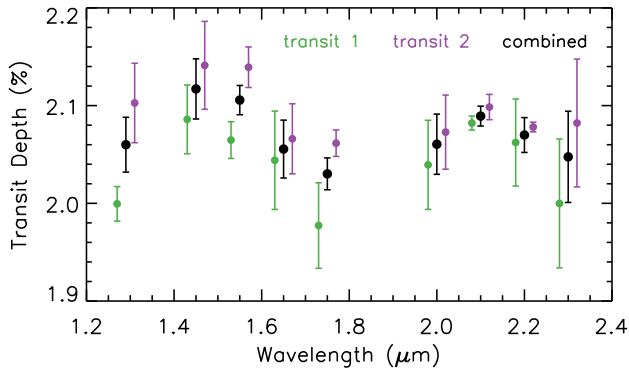}}
\caption{Transit depths with relative errors (i.e. assuming fixed $i$, $b$, and $\gamma_{2}$, see \S3.1.4 ) determined from the analyses of each transit separately and in combination. The transit depths for the individual transits are shown offset in wavelength for clarity.}
\label{fig:transmission_both}
\end{figure}

The transit depths with relative errors (i.e. assuming fixed $i$ and $b$, see \S3.1.4 ) determined from the analyses of each transit separately and in combination are shown in Figure\,\ref{fig:transmission_both}. The depths from the two transit observations all agree within 2.6$\sigma$, and six of the nine values agree within 1.1$\sigma$.

\begin{deluxetable}{cc}
\tabletypesize{\scriptsize}
\tablecolumns{3}
\tablewidth{0pc}
\tablecaption{Secondary Eclipse Depths}
\tablehead{
 \colhead{Wavelength ($\mu$m)} &
 \colhead{F$_{p}$/F$_{\star}$}
}
\startdata
  1.25 -- 1.33 & 0.00083 $\pm$ 3.9E-04 \\
  1.40 -- 1.50 & 0.00208 $\pm$ 4.5E-04 \\
  1.50 -- 1.60 & 0.00180 $\pm$ 1.7E-04 \\
  1.60 -- 1.70 & 0.00200 $\pm$ 3.6E-04 \\
  1.70 -- 1.80 & 0.00188 $\pm$ 3.8E-04 \\
  1.95 -- 2.05 & 0.00238 $\pm$ 3.0E-04 \\
  2.05 -- 2.15 & 0.00227 $\pm$ 1.6E-04 \\
  2.15 -- 2.25 & 0.00242 $\pm$ 3.1E-04 \\
  2.25 -- 2.35 & 0.00312 $\pm$ 9.1E-04
\enddata
\label{tab:eclipse_depths}
\end{deluxetable}

\begin{figure*}
\resizebox{\hsize}{!}{\includegraphics{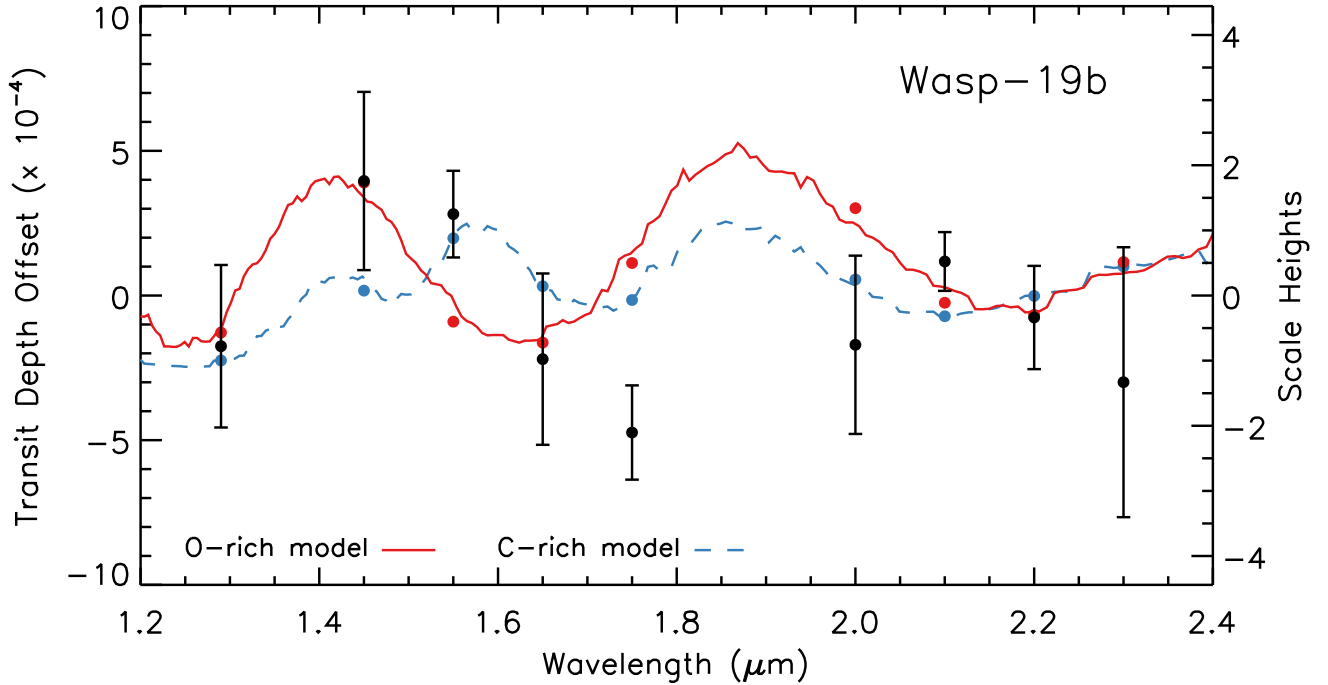}}
\caption{The transmission spectrum of WASP-19b in terms of the relative transit depth (left axis) and scale height of the planet's atmosphere (right axis). The measurements are given as the black circles. Also shown are models (lines) for the planet's transmission spectrum calculated using the methods of \citet{madhusudhan09}. One model has a solar carbon-to-oxygen ratio (``O-rich''; solid red line) and the other model has a carbon-to-oxygen ratio of 1.0 (``C-rich''; dashed blue line). The colored points give the model values binned over the bandpass of the observations. The models have been adjusted up and down to give the best-fit to the data.}
\label{fig:transmission_rel}
\end{figure*}

The transit depths for the combined analysis relative to the average value and in terms of the estimated scale height ($H$) of the atmosphere (546\,km) are shown in Figure\,\ref{fig:transmission_rel}. Two things can be seen from the data without comparison to models. First, the data rule out variations from the mean transmission spectrum of much more than a few scale heights at this resolution. The standard deviation is 1.3\,H, and the maximum spread between points is 3.9\,H. This suggests that low-resolution near-infrared observations of planets like WASP-19b must achieve spectrophotometric precisions on order of one atmospheric scale height or better to confidently detect molecular features.

The second thing that can be seen from the data alone is that a flat line is a poor fit to the measurements. The best-fit flat line gives $\chi^2$\,=\,16.8 for 8 degrees of freedom, which has a 3\% probability to happen by chance. The $K$-band region is featureless within the limits of the precision of the data. This is to be expected given that CH$_{4}$ is the main opacity source at these wavelengths and this molecule is not expected to be present in the atmosphere of such a hot planet like WASP-19b assuming chemical equilibrium holds. However, the $H$-band region of the spectrum exhibits variation.


Theoretical models for WASP-19b's transmission spectrum calculated using the methods of \citet{madhusudhan09} are shown compared to the data in Figure\,\ref{fig:transmission_rel}. We considered two different solar metallicity models for comparison with the data: one with a roughly solar carbon-to-oxygen ratio of 0.5 \citep{asplund09}, and one with a carbon-to-oxygen abundance ratio of 1.0 that was motivated by the recent idea that some hot-Jupiters might have such an abundance pattern \citep{madhusudhan11b,madhusudhan12}. 

The theoretical models are consistent with the data in the sense that variations of only a few scale heights are expected at this resolution. The solar composition model is actually a worse match to the data than the flat line. After adjusting the model by subtracting a fitted offset to account for the unknown pressure level probed by the observations, the solar abundance model gives $\chi^2$\,=\,24.3 for 8 degrees of freedom. This poor fit suggests that the data are inconsistent with the solar abundance model at 3.1$\sigma$ confidence. This inference hinges critically on the points at 1.45 and 1.75\,$\mu$m. The main features in the solar abundance model are due to H$_{2}$O, and these two channels should both show deeper transits compared to the two channels between them if this molecule is present in the planet's atmosphere.

The carbon-rich model, while matching the data better than the solar abundance model, is only a marginally better fit than the flat line with $\chi^2$\,=\,15.4 after determining a best-fit offset. The main spectral features in the carbon-rich model are due to HCN and H$_2$O, though the H$_2$O abundance is lower in the C-rich model than in the O-rich model by over a factor of 10. The molecule HCN has not been detected in an exoplanet atmosphere before, though several theoretical studies have predicted its existence in C-rich atmospheres \citep{madhusudhan11a, kopparapu12, moses13}. Our observations favor the presence of this molecule, but the carbon-rich model does not provide a good global fit. Furthermore, even if a model could be found that provided a good fit to the data, there is only a 2.1$\sigma$ confidence on the detection of spectral features.


\subsection{Emission Spectrum}
\begin{figure}
\resizebox{\hsize}{!}{\includegraphics{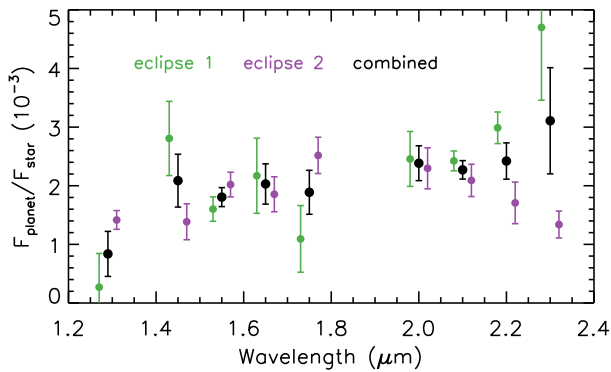}}
\caption{Secondary eclipse depths determined from the analyses of each eclipse separately and in combination. The depths for the individual eclipses are shown offset in wavelength for clarity.}
\label{fig:emission_both}
\end{figure}

The secondary eclipse depths determined from the analyses of each eclipse separately and in combination are shown in Figure\,\ref{fig:emission_both}. The significance of the detections in the combined analysis range from 2.2$\sigma$ (the $J$-band point) to 14.4$\sigma$ (the point at 2.1\,$\mu$m). The data show the characteristic increase in the planet-to-star flux ratio with wavelength that is expected for hot-Jupiter spectra \citep{burrows05, fortney05, seager05}. The depths determined from analyzing the secondary eclipse events separately don't agree as well as the depths determined from analyzing the two primary transit events separately. This is likely due to the difficulty of accurately estimating confidence intervals on low signal-to-noise detections. The eclipse depths from the two different events do all agree within 2.9$\sigma$, and the combined analysis is likely more robust than the analyses of the individual events.

\begin{figure*}
\resizebox{\hsize}{!}{\includegraphics{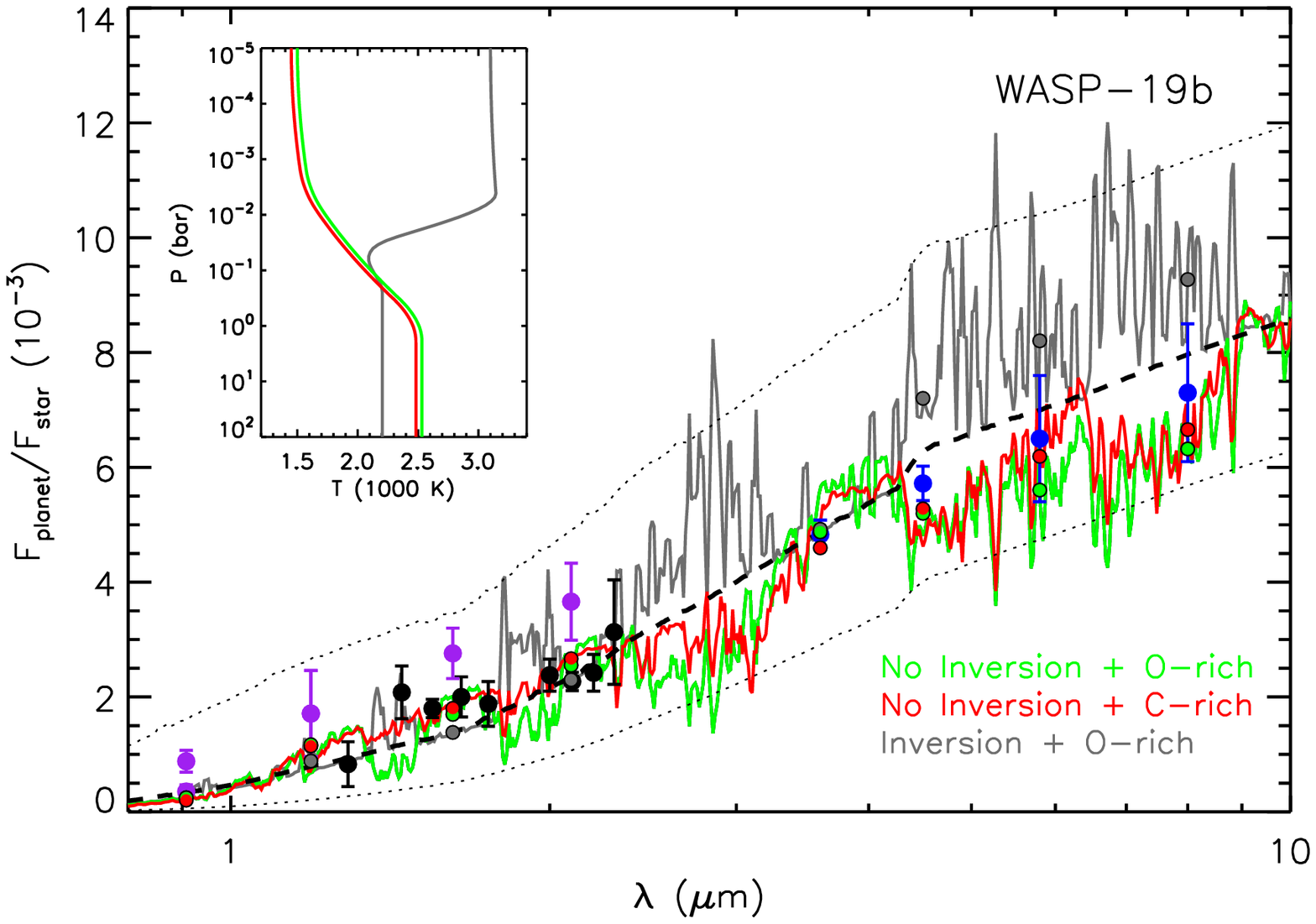}}
\caption{The thermal emission spectrum of WASP-19b in terms of the planet-to-star flux ratio (secondary eclipse depth). The measurements presented in this paper are shown as the black circles. Previous reported measurements from \citet{anderson10}, \citet{gibson10}, \citet{burton12}, \citet{anderson13}, and \citet{lendl13} are also shown as the blue (\textit{Spitzer} IRAC) and purple (ground-based) circles. The green and red lines represent models without thermal inversions that were fit to our data and the \textit{Spitzer} IRAC data using the methods of \citet{madhusudhan09}. The green line is a roughly solar composition model (C/O\,=\,0.5), and the red line is a carbon-rich model (C/O\,=\,1.0). The grey line shows an example model with a thermal inversion and solar composition. The green, red, and grey points show the respective values of the models integrated over the bandpasses of the photometric points. The integrated model values for our spectroscopic data are not shown for clarity. The inset shows the corresponding temperature-pressure profiles for the models. The dashed line is a model of the planet assuming it radiates as a blackbody with a temperature of 2250\,K. The dotted lines are blackbody models for the planet with T\,=\,1800 (lower line) and 2900 (upper line) K.}
\label{fig:emission}
\end{figure*}

The secondary eclipse depths from the combined analysis of the two observed events are shown in Figure\,\ref{fig:emission} along with previously published data and example theoretical models. Previous secondary eclipse measurements of WASP-19b using photometric techniques have been made with with IRAC on \textit{Spitzer} \citep{anderson13}, HAWKI on the VLT \citep{anderson10, gibson10, lendl13}, ULTRACAM on the NTT \cite{burton12}, EulerCam on the Euler-Swiss telescope \citep{lendl13}, and TRAPPIST \citep{lendl13}. Our data overlap broad-band H and narrow-band K measurements previously reported by \citet{anderson10} and \citet{gibson10}, respectively. We performed an analysis of our data with the bandpasses set to closely match the wavelengths of those previous measurements to check for consistency. We find that the results are consistent within 1.7 and 1.3$\sigma$ for the $H$- and $K$-band points, respectively.

The combined data set presented in Figure\,\ref{fig:emission} provides coverage of the planet's spectral energy distribution from 0.9 to 8.0\,$\mu$m that is matched only for a few other exoplanets. We performed a spectral retrieval analysis on the  combination of our data with the IRAC points using the methods of \citet{madhusudhan09} to investigate what constraints can be placed on the composition and structure of the planet's atmosphere.

An acceptable fit to the data can be obtained assuming an isothermal atmosphere with a temperature of about 2250\,K ($\chi^{2}$\,=\,19.2 for 12 degrees of freedom). While a perfectly isothermal atmosphere over the entire day-side of the planet may be unrealistic, an effectively isothermal 1-D temperature-pressure profile could arise if the planet has a thermal inversion where heating due to absorption of incident radiation from the host star causes the temperature to increase at lower pressures instead of decrease \citep{hubeny03, burrows07, knutson08}. The planet's atmosphere could appear isothermal because the heating yields only a modest change in temperature over the pressures probed by the observations. In this case, there could be a large temperature increase at lower pressures that aren't being probed given the resolution and precision of our data. An effectively isothermal atmosphere with a temperature of roughly 3000\,K has also recently been suggested for the planet WASP-12b \citep{crossfield12b}. Higher resolution and higher precision data are needed in both cases to detect spectral features and further constrain the planets' temperature-pressure profiles.

The data for WASP-19b disfavor a thermal inversion where there is a large temperature increase to lower pressures in the observed part of the atmosphere in both oxygen-rich (C/O\,=\,0.5, see Figure \,\ref{fig:emission}) and carbon-rich (C/O\,=\,1.0) composition models. This is consistent with the finding of \citet{anderson13}. In the presence of a strong inversion in the observable atmosphere, absorption lines due to H$_{2}$O and CO that are prominent in the IRAC bands would reverse to emission and yield a poor fit to the \textit{Spitzer} data. 

We also explored models for the planet's atmosphere with no thermal inversions, i.e. where the temperature decreases monotonically with pressure. Figure\,\ref{fig:emission} shows two example models with different chemical compositions: one with an oxygen-rich solar composition (C/O\,=\,0.5) and another with a carbon-rich composition (C/O\,=\,1.0). The C-rich model provides a better fit to the data ($\chi^{2}$\,=\,15.9) compared to the O-rich model ($\chi^2$\,=\,30.7), although the significance is low due to the low number of degrees of freedom, which is only three (the models have 10 free parameters). The absorption features of H$_{2}$O that are expected in the O-rich scenario are not obvious in the MMIRS data, which is consistent with what is seen in the transmission spectrum. On the other hand, though the C-rich model provides an acceptable fit to the data, it is not much better than an isothermal model with any composition given the current precision of the data.

The isothermal and carbon-rich non-inversion models fitted to the IRAC and MMIRS data also provide a reasonable fit to the data at other wavelengths that were not included in the fit. The $z'$-band points of \citet{burton12} and \citet{lendl13} are  inconsistent with each other at 2.4$\sigma$. The fitted models happen to go right through the \citet{lendl13} point. The deeper eclipse observed by \citet{burton12} suggests a higher temperature that is difficult to reconcile with our data.

\section{DISCUSSION}
We have presented the first near-infrared transmission and emission spectroscopy measurements of a hot-Jupiter obtained using the technique of multi-object spectroscopy with wide slits. The data do not yield strong detections of spectral features, yet they do provide some constraints on the physical properties of WASP-19b's atmosphere. We rule out broad spectral features in the planet's near-infrared transmission spectrum that would arise from probing more than a few scale heights in atmospheric pressure. The combination of our thermal emission measurements with extant \textit{Spitzer} IRAC data suggests that the planet does not have a thermal inversion yielding a large increase in temperature to lower pressures in the observed part of its atmosphere. The emission data are consistent with a model for the planet's atmosphere that is isothermal over the observed part of the atmosphere, and with a model that has decreasing temperature with pressure and a C/O value that is substantially larger than the solar value.

A key question raised by our observations is why isn't the obtained precision higher than it is? The current data, and our previous near-infrared observations of GJ\,1214b \citep{bean11}, fall short of delivering light curve residuals better than a factor of two to three times the photon-limited expectations. Furthermore, correlated noise in the light curves degrades the precision of the transit and eclipse depth measurements even more. This is in contrast with observations that we have done at optical wavelengths where we have obtained residuals as small as a few hundred ppm per minute and that are within a few tens of percent of the photon-limited precision \citep[e.g.,][]{bean10}. Our study of the MMIRS detector non-linearity described in \S\,2.2.2 yielded surprising results, and this leads us to believe that the quality of near-infrared detectors could be responsible for the lower than expected precision being obtained at these wavelengths. However, we can't speculate as to what the underlying physical reason might be. We do note that if these observations represent the limit obtainable with ground-based near-infrared spectroscopy, then roughly six transits would have to be observed to detect spectral features in the transmission spectrum of a planet like WASP-19b at better than 5$\sigma$ confidence.

The near-infrared is clearly an important wavelength region for exoplanet atmosphere measurements because of the presence of many different molecular bands. However, ground-based near-infrared spectroscopy to measure these molecules is challenging due to the presence of water vapor in Earth's atmosphere. Numerous and strong telluric water lines sculpt the near-infrared spectrum and limit observations to the canonical $Y$-, $J$-, $H$-, and $K$-bands. Observations at the wavelengths between these bands are simply not possible from the ground because no light from astronomical sources reaches the telescope. The presence of telluric water vapor further complicates matters because it produces lines that even contaminate the windows where light can get through. These lines are expected to be strongly variable over timescales relevant for transit observations due to changing conditions in the atmosphere above an observatory. This is borne out by our observations. We clearly see stronger variations in the atmospheric transparency as measured by the reference stars between the edges and centers of the $H$- and $K$-bands (see Figure\,\ref{fig:decor_lc}).

Variable telluric water vapor has been identified as a possible reason for the strong feature seen in the emission spectrum of the planet HD\,189733b that \citet{swain10} presented \citep{mandell11}. The technique of observing only the transiting planet system, which \citet{swain10} use, requires correcting for telluric variations using the data themselves. The \citet{swain10} approach is to filter out these variations based on the assumption that they are correlated with wavelength and time. In contrast, the multi-object technique we use enables corrections for variations in Earth's atmospheric transparency at every wavelength for every exposure using an external reference. Therefore, our data should be insensitive to the problem of variability in water vapor lines assuming the variations are common mode for sources within a few arcminutes of each other on the sky.



The measurements we have presented here represent one of the few cases of ground-based exoplanet atmosphere observations with repeated observations. The moderate level of disagreement seen between the results from the separate analyses of the different events illustrates the potential pitfalls of making statements about the nature of exoplanetary atmospheres at even a 3$\sigma$ formal confidence level when the inference critically depends on one or two points that are derived from observations of a single event. Our data require little systematic decorrelation and we used a well-regarded algorithm for estimating confidence intervals in the face of correlated noise (residual permutation bootstrap), but yet we still only see repeatability in our transit and eclipse depths at the 2 -- 3$\sigma$ level. One strength of ground-based measurements is that the observations are easier to repeat due to the generally lower time pressure on the telescopes compared to space-based facilities. We want to encourage observers pursuing ground-based observations to take advantage of this opportunity, and we also want to encourage telescope time allocation committees to look favorably on proposed repeated observations not just because they would have a higher formal signal-to-noise, but also because the results will likely be more robust.

\acknowledgments
J.L.B. acknowledges support from the Alfred P.~Sloan Foundation. J.-M.D. acknowledges funding from NASA through the Sagan Exoplanet Fellowship program administered by the NASA Exoplanet Science Institute (NExScI). The results presented are based on observations made with the 6.5\,m Magellan telescopes located at Las Campanas Observatory.

{\it Facilities:} \facility{Magellan:Clay (MMIRS)}


\begin{thebibliography}{70}
\expandafter\ifx\csname natexlab\endcsname\relax\def\natexlab#1{#1}\fi

\bibitem[{{Alonso} {et~al.}(2010){Alonso}, {Deeg}, {Kabath}, \&
  {Rabus}}]{alonso10}
{Alonso}, R., {Deeg}, H.~J., {Kabath}, P., \& {Rabus}, M. 2010, \aj, 139, 1481

\bibitem[{{Anderson} {et~al.}(2010){Anderson}, {Gillon}, {Maxted}, {Barman},
  {Collier Cameron}, {Hellier}, {Queloz}, {Smalley}, \& {Triaud}}]{anderson10}
{Anderson}, D.~R., {et~al.} 2010, \aap, 513, L3+

\bibitem[{{Anderson} {et~al.}(2013){Anderson}, {Smith}, {Madhusudhan},
  {Wheatley}, {Collier Cameron}, {Hellier}, {Campo}, {Gillon}, {Harrington},
  {Maxted}, {Pollacco}, {Queloz}, {Smalley}, {Triaud}, \& {West}}]{anderson13}
---. 2013, \mnras, 430, 3422

\bibitem[{{Asplund} {et~al.}(2009){Asplund}, {Grevesse}, {Sauval}, \&
  {Scott}}]{asplund09}
{Asplund}, M., {Grevesse}, N., {Sauval}, A.~J., \& {Scott}, P. 2009, \araa, 47,
  481

\bibitem[{{Bean} {et~al.}(2010){Bean}, {Miller-Ricci Kempton}, \&
  {Homeier}}]{bean10}
{Bean}, J.~L., {Miller-Ricci Kempton}, E., \& {Homeier}, D. 2010, \nat, 468,
  669

\bibitem[{{Bean} {et~al.}(2011){Bean}, {D{\'e}sert}, {Kabath}, {Stalder},
  {Seager}, {Miller-Ricci Kempton}, {Berta}, {Homeier}, {Walsh}, \&
  {Seifahrt}}]{bean11}
{Bean}, J.~L., {et~al.} 2011, \apj, 743, 92

\bibitem[{{Berta} {et~al.}(2011){Berta}, {Charbonneau}, {Bean}, {Irwin},
  {Burke}, {D{\'e}sert}, {Nutzman}, \& {Falco}}]{berta11}
{Berta}, Z.~K., {Charbonneau}, D., {Bean}, J., {Irwin}, J., {Burke}, C.~J.,
  {D{\'e}sert}, J.-M., {Nutzman}, P., \& {Falco}, E.~E. 2011, \apj, 736, 12

\bibitem[{{Budaj}(2011)}]{budaj11}
{Budaj}, J. 2011, \aj, 141, 59

\bibitem[{{Burrows} {et~al.}(2008){Burrows}, {Budaj}, \& {Hubeny}}]{burrows08}
{Burrows}, A., {Budaj}, J., \& {Hubeny}, I. 2008, \apj, 678, 1436

\bibitem[{{Burrows} {et~al.}(2007){Burrows}, {Hubeny}, {Budaj}, {Knutson}, \&
  {Charbonneau}}]{burrows07}
{Burrows}, A., {Hubeny}, I., {Budaj}, J., {Knutson}, H.~A., \& {Charbonneau},
  D. 2007, \apjl, 668, L171

\bibitem[{{Burrows} {et~al.}(2005){Burrows}, {Hubeny}, \&
  {Sudarsky}}]{burrows05}
{Burrows}, A., {Hubeny}, I., \& {Sudarsky}, D. 2005, \apjl, 625, L135

\bibitem[{{Burton} {et~al.}(2012){Burton}, {Watson}, {Littlefair}, {Dhillon},
  {Gibson}, {Marsh}, \& {Pollacco}}]{burton12}
{Burton}, J.~R., {Watson}, C.~A., {Littlefair}, S.~P., {Dhillon}, V.~S.,
  {Gibson}, N.~P., {Marsh}, T.~R., \& {Pollacco}, D. 2012, \apjs, 201, 36

\bibitem[{{Cowan} \& {Agol}(2011)}]{cowan11}
{Cowan}, N.~B., \& {Agol}, E. 2011, \apj, 729, 54

\bibitem[{{Croll} {et~al.}(2011{\natexlab{a}}){Croll}, {Albert},
  {Jayawardhana}, {Miller-Ricci Kempton}, {Fortney}, {Murray}, \&
  {Neilson}}]{croll11a}
{Croll}, B., {Albert}, L., {Jayawardhana}, R., {Miller-Ricci Kempton}, E.,
  {Fortney}, J.~J., {Murray}, N., \& {Neilson}, H. 2011{\natexlab{a}}, \apj,
  736, 78

\bibitem[{{Croll} {et~al.}(2010{\natexlab{a}}){Croll}, {Albert}, {Lafreniere},
  {Jayawardhana}, \& {Fortney}}]{croll10a}
{Croll}, B., {Albert}, L., {Lafreniere}, D., {Jayawardhana}, R., \& {Fortney},
  J.~J. 2010{\natexlab{a}}, \apj, 717, 1084

\bibitem[{{Croll} {et~al.}(2010{\natexlab{b}}){Croll}, {Jayawardhana},
  {Fortney}, {Lafreni{\`e}re}, \& {Albert}}]{croll10b}
{Croll}, B., {Jayawardhana}, R., {Fortney}, J.~J., {Lafreni{\`e}re}, D., \&
  {Albert}, L. 2010{\natexlab{b}}, \apj, 718, 920

\bibitem[{{Croll} {et~al.}(2011{\natexlab{b}}){Croll}, {Lafreniere}, {Albert},
  {Jayawardhana}, {Fortney}, \& {Murray}}]{croll11b}
{Croll}, B., {Lafreniere}, D., {Albert}, L., {Jayawardhana}, R., {Fortney},
  J.~J., \& {Murray}, N. 2011{\natexlab{b}}, \aj, 141, 30

\bibitem[{{Crossfield} {et~al.}(2011){Crossfield}, {Barman}, \&
  {Hansen}}]{crossfield11}
{Crossfield}, I.~J.~M., {Barman}, T., \& {Hansen}, B.~M.~S. 2011, \apj, 736,
  132

\bibitem[{{Crossfield} {et~al.}(2012{\natexlab{a}}){Crossfield}, {Barman},
  {Hansen}, {Tanaka}, \& {Kodama}}]{crossfield12b}
{Crossfield}, I.~J.~M., {Barman}, T., {Hansen}, B.~M.~S., {Tanaka}, I., \&
  {Kodama}, T. 2012{\natexlab{a}}, \apj, 760, 140

\bibitem[{{Crossfield} {et~al.}(2012{\natexlab{b}}){Crossfield}, {Hansen}, \&
  {Barman}}]{crossfield12a}
{Crossfield}, I.~J.~M., {Hansen}, B.~M.~S., \& {Barman}, T. 2012{\natexlab{b}},
  \apj, 746, 46

\bibitem[{{de Mooij} {et~al.}(2011){de Mooij}, {de Kok}, {Nefs}, \&
  {Snellen}}]{demooij11}
{de Mooij}, E.~J.~W., {de Kok}, R.~J., {Nefs}, S.~V., \& {Snellen}, I.~A.~G.
  2011, \aap, 528, A49

\bibitem[{{de Mooij} \& {Snellen}(2009)}]{demooij09}
{de Mooij}, E.~J.~W., \& {Snellen}, I.~A.~G. 2009, \aap, 493, L35

\bibitem[{{de Mooij} {et~al.}(2012){de Mooij}, {Brogi}, {de Kok},
  {Koppenhoefer}, {Nefs}, {Snellen}, {Greiner}, {Hanse}, {Heinsbroek}, {Lee},
  \& {van der Werf}}]{demooij12}
{de Mooij}, E.~J.~W., {et~al.} 2012, \aap, 538, A46

\bibitem[{{Deming} {et~al.}(2012){Deming}, {Fraine}, {Sada}, {Madhusudhan},
  {Knutson}, {Harrington}, {Blecic}, {Nymeyer}, {Smith}, \&
  {Jackson}}]{deming12}
{Deming}, D., {et~al.} 2012, \apj, 754, 106

\bibitem[{{D{\'e}sert} {et~al.}(2011){D{\'e}sert}, {Sing}, {Vidal-Madjar},
  {H{\'e}brard}, {Ehrenreich}, {Lecavelier Des Etangs}, {Parmentier}, {Ferlet},
  \& {Henry}}]{desert11}
{D{\'e}sert}, J.-M., {et~al.} 2011, \aap, 526, A12

\bibitem[{{Doyle} {et~al.}(2013){Doyle}, {Smalley}, {Maxted}, {Anderson},
  {Cameron}, {Gillon}, {Hellier}, {Pollacco}, {Queloz}, {Triaud}, \&
  {West}}]{doyle13}
{Doyle}, A.~P., {et~al.} 2013, \mnras, 428, 3164

\bibitem[{{Dragomir} {et~al.}(2011){Dragomir}, {Kane}, {Pilyavsky},
  {Mahadevan}, {Ciardi}, {Gazak}, {Gelino}, {Payne}, {Rabus}, {Ramirez}, {von
  Braun}, {Wright}, \& {Wyatt}}]{dragomir11}
{Dragomir}, D., {et~al.} 2011, \aj, 142, 115

\bibitem[{{Eastman} {et~al.}(2010){Eastman}, {Siverd}, \& {Gaudi}}]{eastman10}
{Eastman}, J., {Siverd}, R., \& {Gaudi}, B.~S. 2010, \pasp, 122, 935

\bibitem[{{Fortney} {et~al.}(2008){Fortney}, {Lodders}, {Marley}, \&
  {Freedman}}]{fortney08}
{Fortney}, J.~J., {Lodders}, K., {Marley}, M.~S., \& {Freedman}, R.~S. 2008,
  \apj, 678, 1419

\bibitem[{{Fortney} {et~al.}(2005){Fortney}, {Marley}, {Lodders}, {Saumon}, \&
  {Freedman}}]{fortney05}
{Fortney}, J.~J., {Marley}, M.~S., {Lodders}, K., {Saumon}, D., \& {Freedman},
  R. 2005, \apjl, 627, L69

\bibitem[{{Gibson} {et~al.}(2013){Gibson}, {Aigrain}, {Barstow}, {Evans},
  {Fletcher}, \& {Irwin}}]{gibson13}
{Gibson}, N.~P., {Aigrain}, S., {Barstow}, J.~K., {Evans}, T.~M., {Fletcher},
  L.~N., \& {Irwin}, P.~G.~J. 2013, \mnras, 428, 3680

\bibitem[{{Gibson} {et~al.}(2010){Gibson}, {Aigrain}, {Pollacco}, {Barros},
  {Hebb}, {Hrudkov{\'a}}, {Simpson}, {Skillen}, \& {West}}]{gibson10}
{Gibson}, N.~P., {et~al.} 2010, \mnras, 404, L114

\bibitem[{{Gillon} {et~al.}(2009){Gillon}, {Demory}, {Triaud}, {Barman},
  {Hebb}, {Montalb{\'a}n}, {Maxted}, {Queloz}, {Deleuil}, \&
  {Magain}}]{gillon09}
{Gillon}, M., {et~al.} 2009, \aap, 506, 359

\bibitem[{{Hauschildt} {et~al.}(1999){Hauschildt}, {Allard}, {Ferguson},
  {Baron}, \& {Alexander}}]{hauschildt99}
{Hauschildt}, P.~H., {Allard}, F., {Ferguson}, J., {Baron}, E., \& {Alexander},
  D.~R. 1999, \apj, 525, 871

\bibitem[{{Hayek} {et~al.}(2012){Hayek}, {Sing}, {Pont}, \&
  {Asplund}}]{hayek12}
{Hayek}, W., {Sing}, D., {Pont}, F., \& {Asplund}, M. 2012, \aap, 539, A102

\bibitem[{{Hebb} {et~al.}(2010){Hebb}, {Collier-Cameron}, {Triaud}, {Lister},
  {Smalley}, {Maxted}, {Hellier}, {Anderson}, {Pollacco}, {Gillon}, {Queloz},
  {West}, {Bentley}, {Enoch}, {Haswell}, {Horne}, {Mayor}, {Pepe}, {Segransan},
  {Skillen}, {Udry}, \& {Wheatley}}]{hebb10}
{Hebb}, L., {et~al.} 2010, \apj, 708, 224

\bibitem[{{Hellier} {et~al.}(2011){Hellier}, {Anderson}, {Collier-Cameron},
  {Miller}, {Queloz}, {Smalley}, {Southworth}, \& {Triaud}}]{hellier11}
{Hellier}, C., {Anderson}, D.~R., {Collier-Cameron}, A., {Miller}, G.~R.~M.,
  {Queloz}, D., {Smalley}, B., {Southworth}, J., \& {Triaud}, A.~H.~M.~J. 2011,
  \apjl, 730, L31

\bibitem[{{Hubeny} {et~al.}(2003){Hubeny}, {Burrows}, \& {Sudarsky}}]{hubeny03}
{Hubeny}, I., {Burrows}, A., \& {Sudarsky}, D. 2003, \apj, 594, 1011

\bibitem[{{Knutson} {et~al.}(2008){Knutson}, {Charbonneau}, {Allen}, {Burrows},
  \& {Megeath}}]{knutson08}
{Knutson}, H.~A., {Charbonneau}, D., {Allen}, L.~E., {Burrows}, A., \&
  {Megeath}, S.~T. 2008, \apj, 673, 526

\bibitem[{{Knutson} {et~al.}(2007){Knutson}, {Charbonneau}, {Noyes}, {Brown},
  \& {Gilliland}}]{knutson07}
{Knutson}, H.~A., {Charbonneau}, D., {Noyes}, R.~W., {Brown}, T.~M., \&
  {Gilliland}, R.~L. 2007, \apj, 655, 564

\bibitem[{{Knutson} {et~al.}(2010){Knutson}, {Howard}, \&
  {Isaacson}}]{knutson10}
{Knutson}, H.~A., {Howard}, A.~W., \& {Isaacson}, H. 2010, \apj, 720, 1569

\bibitem[{{Kopparapu} {et~al.}(2012){Kopparapu}, {Kasting}, \&
  {Zahnle}}]{kopparapu12}
{Kopparapu}, R.~k., {Kasting}, J.~F., \& {Zahnle}, K.~J. 2012, \apj, 745, 77

\bibitem[{{Kulas} {et~al.}(2012){Kulas}, {McLean}, \& {Steidel}}]{kulas12}
{Kulas}, K.~R., {McLean}, I.~S., \& {Steidel}, C.~C. 2012, in Society of
  Photo-Optical Instrumentation Engineers (SPIE) Conference Series, Vol. 8453,
  Society of Photo-Optical Instrumentation Engineers (SPIE) Conference Series

\bibitem[{{Lendl} {et~al.}(2013){Lendl}, {Gillon}, {Queloz}, {Alonso}, {Fumel},
  {Jehin}, \& {Naef}}]{lendl13}
{Lendl}, M., {Gillon}, M., {Queloz}, D., {Alonso}, R., {Fumel}, A., {Jehin},
  E., \& {Naef}, D. 2013, \aap, 552, A2

\bibitem[{{L{\'o}pez-Morales} {et~al.}(2010){L{\'o}pez-Morales}, {Coughlin},
  {Sing}, {Burrows}, {Apai}, {Rogers}, {Spiegel}, \& {Adams}}]{lopezmorales10}
{L{\'o}pez-Morales}, M., {Coughlin}, J.~L., {Sing}, D.~K., {Burrows}, A.,
  {Apai}, D., {Rogers}, J.~C., {Spiegel}, D.~S., \& {Adams}, E.~R. 2010, \apjl,
  716, L36

\bibitem[{{Madhusudhan}(2012)}]{madhusudhan12}
{Madhusudhan}, N. 2012, \apj, 758, 36

\bibitem[{{Madhusudhan} {et~al.}(2011{\natexlab{a}}){Madhusudhan}, {Mousis},
  {Johnson}, \& {Lunine}}]{madhusudhan11a}
{Madhusudhan}, N., {Mousis}, O., {Johnson}, T.~V., \& {Lunine}, J.~I.
  2011{\natexlab{a}}, \apj, 743, 191

\bibitem[{{Madhusudhan} \& {Seager}(2009)}]{madhusudhan09}
{Madhusudhan}, N., \& {Seager}, S. 2009, \apj, 707, 24

\bibitem[{{Madhusudhan} {et~al.}(2011{\natexlab{b}}){Madhusudhan},
  {Harrington}, {Stevenson}, {Nymeyer}, {Campo}, {Wheatley}, {Deming},
  {Blecic}, {Hardy}, {Lust}, {Anderson}, {Collier-Cameron}, {Britt}, {Bowman},
  {Hebb}, {Hellier}, {Maxted}, {Pollacco}, \& {West}}]{madhusudhan11b}
{Madhusudhan}, N., {et~al.} 2011{\natexlab{b}}, \nat, 469, 64

\bibitem[{{Mandel} \& {Agol}(2002)}]{mandel02}
{Mandel}, K., \& {Agol}, E. 2002, \apjl, 580, L171

\bibitem[{{Mandell} {et~al.}(2011){Mandell}, {Drake Deming}, {Blake},
  {Knutson}, {Mumma}, {Villanueva}, \& {Salyk}}]{mandell11}
{Mandell}, A.~M., {Drake Deming}, L., {Blake}, G.~A., {Knutson}, H.~A.,
  {Mumma}, M.~J., {Villanueva}, G.~L., \& {Salyk}, C. 2011, \apj, 728, 18

\bibitem[{{Markwardt}(2009)}]{markwardt09}
{Markwardt}, C.~B. 2009, in Astronomical Society of the Pacific Conference
  Series, Vol. 411, Astronomical Data Analysis Software and Systems XVIII, ed.
  {D.~A.~Bohlender, D.~Durand, \& P.~Dowler}, 251--+

\bibitem[{{McLean} {et~al.}(2010){McLean}, {Steidel}, {Epps}, {Matthews},
  {Adkins}, {Konidaris}, {Weber}, {Aliado}, {Brims}, {Canfield}, {Cromer},
  {Fucik}, {Kulas}, {Mace}, {Magnone}, {Rodriguez}, {Wang}, \&
  {Weiss}}]{mclean10}
{McLean}, I.~S., {et~al.} 2010, in Society of Photo-Optical Instrumentation
  Engineers (SPIE) Conference Series, Vol. 7735, Society of Photo-Optical
  Instrumentation Engineers (SPIE) Conference Series

\bibitem[{{McLeod} {et~al.}(2012){McLeod}, {Fabricant}, {Nystrom}, {McCracken},
  {Amato}, {Bergner}, {Brown}, {Burke}, {Chilingarian}, {Conroy}, {Curley},
  {Furesz}, {Geary}, {Hertz}, {Holwell}, {Matthews}, {Norton}, {Park}, {Roll},
  {Zajac}, {Epps}, \& {Martini}}]{mcleod12}
{McLeod}, B., {et~al.} 2012, \pasp, 124, 1318

\bibitem[{{Moses} {et~al.}(2013){Moses}, {Madhusudhan}, {Visscher}, \&
  {Freedman}}]{moses13}
{Moses}, J.~I., {Madhusudhan}, N., {Visscher}, C., \& {Freedman}, R.~S. 2013,
  \apj, 763, 25

\bibitem[{{Murgas} {et~al.}(2012){Murgas}, {Pall{\'e}}, {Cabrera-Lavers},
  {Col{\'o}n}, {Mart{\'{\i}}n}, \& {Parviainen}}]{murgas12}
{Murgas}, F., {Pall{\'e}}, E., {Cabrera-Lavers}, A., {Col{\'o}n}, K.~D.,
  {Mart{\'{\i}}n}, E.~L., \& {Parviainen}, H. 2012, \aap, 544, A41

\bibitem[{{Narita} {et~al.}(2012){Narita}, {Nagayama}, {Suenaga}, {Fukui},
  {Ikoma}, {Nakajima}, {Nishiyama}, \& {Tamura}}]{narita12}
{Narita}, N., {Nagayama}, T., {Suenaga}, T., {Fukui}, A., {Ikoma}, M.,
  {Nakajima}, Y., {Nishiyama}, S., \& {Tamura}, M. 2012, PASJ in press,
  arXiv:1210.3169

\bibitem[{{Pollacco} {et~al.}(2006){Pollacco}, {Skillen}, {Collier Cameron},
  {Christian}, {Hellier}, {Irwin}, {Lister}, {Street}, {West}, {Anderson},
  {Clarkson}, {Deeg}, {Enoch}, {Evans}, {Fitzsimmons}, {Haswell}, {Hodgkin},
  {Horne}, {Kane}, {Keenan}, {Maxted}, {Norton}, {Osborne}, {Parley}, {Ryans},
  {Smalley}, {Wheatley}, \& {Wilson}}]{pollacco06}
{Pollacco}, D.~L., {et~al.} 2006, \pasp, 118, 1407

\bibitem[{{Redfield} {et~al.}(2008){Redfield}, {Endl}, {Cochran}, \&
  {Koesterke}}]{redfield08}
{Redfield}, S., {Endl}, M., {Cochran}, W.~D., \& {Koesterke}, L. 2008, \apjl,
  673, L87

\bibitem[{{Rogers} {et~al.}(2009){Rogers}, {Apai}, {L{\'o}pez-Morales}, {Sing},
  \& {Burrows}}]{rogers09}
{Rogers}, J.~C., {Apai}, D., {L{\'o}pez-Morales}, M., {Sing}, D.~K., \&
  {Burrows}, A. 2009, \apj, 707, 1707

\bibitem[{{Sada} {et~al.}(2010){Sada}, {Deming}, {Jackson}, {Jennings},
  {Peterson}, {Haase}, {Bays}, {O'Gorman}, \& {Lundsford}}]{sada10}
{Sada}, P.~V., {et~al.} 2010, \apjl, 720, L215

\bibitem[{{Seager} {et~al.}(2005){Seager}, {Richardson}, {Hansen}, {Menou},
  {Cho}, \& {Deming}}]{seager05}
{Seager}, S., {Richardson}, L.~J., {Hansen}, B.~M.~S., {Menou}, K., {Cho},
  J.~Y.-K., \& {Deming}, D. 2005, \apj, 632, 1122

\bibitem[{{Sing} \& {L{\'o}pez-Morales}(2009)}]{sing09}
{Sing}, D.~K., \& {L{\'o}pez-Morales}, M. 2009, \aap, 493, L31

\bibitem[{{Smith} {et~al.}(2011){Smith}, {Anderson}, {Skillen}, {Collier
  Cameron}, \& {Smalley}}]{smith11}
{Smith}, A.~M.~S., {Anderson}, D.~R., {Skillen}, I., {Collier Cameron}, A., \&
  {Smalley}, B. 2011, \mnras, 416, 2096

\bibitem[{{Snellen} {et~al.}(2008){Snellen}, {Albrecht}, {de Mooij}, \& {Le
  Poole}}]{snellen08}
{Snellen}, I.~A.~G., {Albrecht}, S., {de Mooij}, E.~J.~W., \& {Le Poole}, R.~S.
  2008, \aap, 487, 357

\bibitem[{{Swain} {et~al.}(2010){Swain}, {Deroo}, {Griffith}, {Tinetti},
  {Thatte}, {Vasisht}, {Chen}, {Bouwman}, {Crossfield}, {Angerhausen},
  {Afonso}, \& {Henning}}]{swain10}
{Swain}, M.~R., {et~al.} 2010, \nat, 463, 637

\bibitem[{{Tregloan-Reed} {et~al.}(2013){Tregloan-Reed}, {Southworth}, \&
  {Tappert}}]{tregloan13}
{Tregloan-Reed}, J., {Southworth}, J., \& {Tappert}, C. 2013, \mnras, 428, 3671

\bibitem[{{Waldmann} {et~al.}(2012){Waldmann}, {Tinetti}, {Drossart}, {Swain},
  {Deroo}, \& {Griffith}}]{waldmann12}
{Waldmann}, I.~P., {Tinetti}, G., {Drossart}, P., {Swain}, M.~R., {Deroo}, P.,
  \& {Griffith}, C.~A. 2012, \apj, 744, 35

\bibitem[{{Zhao} {et~al.}(2012{\natexlab{a}}){Zhao}, {Milburn}, {Barman},
  {Hinkley}, {Swain}, {Wright}, \& {Monnier}}]{zhao12b}
{Zhao}, M., {Milburn}, J., {Barman}, T., {Hinkley}, S., {Swain}, M.~R.,
  {Wright}, J., \& {Monnier}, J.~D. 2012{\natexlab{a}}, \apjl, 748, L8

\bibitem[{{Zhao} {et~al.}(2012{\natexlab{b}}){Zhao}, {Monnier}, {Swain},
  {Barman}, \& {Hinkley}}]{zhao12a}
{Zhao}, M., {Monnier}, J.~D., {Swain}, M.~R., {Barman}, T., \& {Hinkley}, S.
  2012{\natexlab{b}}, \apj, 744, 122

\end{thebibliography}
\end{document}